\documentstyle[prd,eqsecnum,aps]{revtex}

\def\be{\begin{equation}}
\def\ee{\end{equation}}
\def\ba{\begin{eqnarray}}
\def\ea{\end{eqnarray}}

\def\negenspace{\kern-1.1em}

\begin{document}
\draft

\twocolumn

\title{Comment on: Topological invariants, instantons, and the chiral anomaly on
spaces with torsion}

\author{Dirk Kreimer$^\$ $\thanks{E-mail: kreimer@thep.physik.uni-mainz.de} and
Eckehard W. Mielke$^\diamond$\thanks{E-mail: ekke@xanum.uam.mx} \\
$^\$ $ Institut f\"ur Physik, Universit\"at Mainz\\
55099 Mainz, GERMANY\\
$^{\diamond}$ Departamento de F\'{\i}sica,
Universidad Aut\'onoma Metropolitana--Iztapalapa,\\
Apartado Postal 55-534, C.P. 09340, M\'exico, D.F., MEXICO\\}

\maketitle

\today

\begin{abstract}
In Riemann--Cartan spacetimes
with torsion only its axial covector piece  $A$
couples to {\em massive} Dirac fields. Using renormalization group
arguments, we show that
besides  the familiar Riemannian  term only the Pontrjagin type four--form
$dA\wedge dA$ does arise additionally in the {\em chiral anomaly}, 
but {\em not} the Nieh--Yan term  $d\,^*\! A$, as has been  claimed  in a recent paper
[PRD {\bf 55}, 7580 (1997)].

\end{abstract}
\section{Introduction}
Quantum anomalies both in the Riemannian and in the
Riemann-Cartan spacetimes were calculated previously
using different methods,
see e.g. \cite{torsfree,Yajima}. However,
recently \cite{ChandiaZ}
the completeness of these earlier calculations have been questioned
which all demonstrated that the Nieh--Yan four-form \cite{NY} is irrelevant to the axial anomaly.

For the axial anomaly, we have a couple of distinguished
features. Most prominent is its relation with the Atiyah--Singer
index theorem. But also from the viewpoint of perturbative
{\em quantum} field theory (QFT), the chiral anomaly has
some features which signal its conceptual importance.
For all topological field theories like
BF-theories, Chern--Simons, and for all
topological effects like the anomaly,
the remarkable fact holds that the relevant
invariants  do not renormalize --- higher
order loop corrections do not alter the one-loop value
of the anomaly, for example. The fact
that the anomaly is stable against radiative corrections
guarantees that it can be given a topological interpretation.
For the anomaly,
this is the Adler--Bardeen theorem, while other topological field
theories are carefully designed  to have,
amongst other properties, vanishing beta functions.
Another feature is finiteness: in any approach, the chiral anomaly
as a topological invariant is a finite quantity.

In a spacetime with torsion,
Chandia and Zanelli \cite{ChandiaZ}  argue that the Nieh--Yan (NY)
four-form $d\,^*\! A$  will add to this
quantity. As usual, they confront the fact that such a term,
if it is generated at all, is ill-defined, independent of the
regularization. In their case, they use a Fujikawa-type approach
and propose to absorb the regulator mass in a rescaled vierbein.

However, there is a severe misunderstanding in the Ref. \cite{ChandiaZ}. While there is no doubt that
the NY  term can be possibly generated, as demonstrated previously
\cite{Yuri1,Yajima},
this is not the end of the argument.   In order
to obtain  a finite quantity, the tetrads have to be rescaled. While
this might look as an innocent manipulation, this is not so.
In rescaling the  tetrad, the authors of Ref. \cite{ChandiaZ}  ignore
the presence of renormalization conditions and the generation
of a scale upon renormalization. Rescaling the tetrad would
ultimately change the wave function renormalization $Z$-factor.

This factor creeps into the definition of the NY  term at the
quantum level, and thus a rescaling of the tetrad does not
achieve the desired goals. This is not to be surprised: QFT
demands a new $Z$-factor for the NY  term,
in sharp contrast to proper topological invariants at the
quantum level, which remain unchanged under renormalization.

With no renormalization condition available for the NY  term,
and other methods obtaining it as zero, we can only conclude that
the response function of the quantum field theory to a gauge
variation (this is the anomaly) delivers no NY  term.
Or, saying it differently, its finite value is zero after
renormalization.

\section{Gravitational Chern--Simons and Pontrjagin terms}

In our notation,  Clifford--algebra valued
exterior forms \cite{MMM96},
the constant Dirac matrices $\gamma_\alpha$ obeying
$\gamma_\alpha\,\gamma_\beta +\gamma_\beta\gamma_\alpha=2o_{\alpha\beta}$
are saturating the index of the orthonomal coframe one--form $\vartheta^\alpha$ and its
Hodge dual $\eta^\alpha:={}^\ast\vartheta^\alpha$ via
$\gamma:=\gamma_\alpha\vartheta^\alpha$ and
${}^\ast\gamma=\gamma^\alpha\eta_\alpha\,.$
In terms of the  {\em connection}
$\Gamma := {i\over 4} \Gamma^{\alpha\beta}\,\sigma_{\alpha\beta}$, the
$SL(2,C)$--covariant
exterior derivative is given by $D=d+ \Gamma\wedge$,
where
${\sigma}_{\alpha\beta}= \frac{i}{2}
(\gamma_\alpha\gamma_\beta-\gamma_\beta\gamma_\alpha)$ are
the Lorentz generators entering also in the
Clifford-algebra valued two-form
$\sigma:={i\over 2}\gamma\wedge\gamma = {1\over 2}\,{\sigma}_{\alpha\beta}
\,\vartheta^\alpha\wedge\vartheta^\beta$.

Differentiation of these independent
variables leads to the Clifford
algebra--valued two--forms of {\em torsion}
$\Theta :=D\gamma =T^{\alpha}\gamma_{\alpha}$
and {\em curvature}
$\Omega := d\Gamma +\Gamma\wedge \Gamma =
{i\over 4}R^{\alpha\beta}\,\sigma_{\alpha\beta}$
of Riemann--Cartan (RC) geometry.

The {\em Chern--Simons} (CS) term \cite{PRs} for the  Lorentz
connection $C_{\rm RR}  :=$ $- Tr\, \big( {\Gamma}\wedge {\Omega}
- {1\over 3} {\Gamma}\wedge {\Gamma}\wedge  {\Gamma}\big)$  and
its corresponding Pontrjagin term
 $dC_{\rm RR}  = - Tr\, \left( {\Omega}\wedge {\Omega}\right)=
{1\over 2}\,R^{\alpha\beta} \wedge R_{\alpha\beta}$ have the familiar form.
Since the coframe is the translational part of the
Cartan connection \cite{PRs}, there arises also the
{\em translational} CS term \cite{Mi92}
\begin{equation}
C_{\rm TT}  :=  {1\over{8\ell^2}} Tr\, ( {\gamma} \wedge  {\Theta} )=
\frac{1}{2}
\left( C_{\rm RR} -\hat C_{\rm RR}\right)
\label{CTT}
\end{equation}
which is related to the  Nieh--Yan four--form \cite{NY}:
\begin{equation}
dC_{\rm TT}
={2\over{\ell^2}} \left(T^\alpha\wedge
T_\alpha+R_{\alpha\beta}\wedge\vartheta^\alpha\wedge\vartheta^\beta\right) \,.
\label{eq:NY}
\end{equation}

A fundamental length $\ell$ unavoidably occurs here for
dimensional reasons. This can be also motivated
by a de Sitter type \cite{GH} approach, in which the $sl(5,R)$--valued
 connection   $\hat\Gamma =\Gamma +(1/\ell)(\vartheta^\alpha L^4{}_\alpha +
 \vartheta_\beta L^\beta{}_4{})$ is expanded into the dimensionless linear connection
 $\Gamma$ plus the coframe $\vartheta^\alpha= e_i{}^\alpha\, dx^i$
 with canonical dimension $[length]$.
The corresponding CS term $\hat C_{\rm RR}$
splits via
$\hat C_{\rm RR} =C_{\rm RR} -2 C_{\rm TT}$
into the linear one
and that of translations, see
the footnote 31 of Ref. \cite{PRs}. This relation has recently been
``recovered" by Chandia and Zanelli \cite{ChandiaZ}.

The one--form of {\em axial vector torsion}
\be
A:={1\over 4}\,Tr\left(\check{\gamma}\rfloor
{^*\Theta}\right)= {1\over 4}\,^*Tr(\gamma\wedge\Theta) =
\,^*(\vartheta^\alpha\wedge T_\alpha) \label{axitor}
\ee
is a {\em conformal invariant} under the combined transformation of
{\em classical} Weyl rescalings of the coframe, in contrast to
$\,^*\!A=-2\ell^2C_{\rm TT}$, cf. Eqs. (3.14.1,9) of Ref. \cite{PRs}.

\section{Dirac fields in Riemann--Cartan spacetime}
The Dirac Lagrangian  is given by the manifestly {\em Hermitian}
four--form
\ba
 L_{\rm D}(\gamma,\psi,D\psi)&=&
  {i\over 2}\left\{\overline{\psi}\,{^*\gamma}\wedge D\psi
  +\overline{D\psi}\wedge{^*\gamma}\,\psi\right\}+{^* m}\,
\overline{\psi}\psi\nonumber\\
&=&L(\gamma,\psi,D^{\{\}}\psi)
  -{1\over 4}\,A  \wedge\overline{\psi}\gamma_5\,^* \gamma\psi\,,
\label{decldirac}
\ea
for which $\overline{\psi}:=\psi^\dagger\gamma_0$ is the Dirac adjoint
and $^* m=m\eta$ the mass term, cf. \cite{MMM96}.

The  decomposed Lagrangian (\ref{decldirac}) leads to the
following  form of the Dirac equation
\begin{equation}
i \,^*\gamma\wedge \stackrel{\smile}{D}\psi + \,^*m\psi =
i \,^*\gamma\wedge \left[D^{\{\}}  +{i\over 4}\,m\gamma +
{i\over 4}\, A\gamma_5\right]\psi= 0
\label{tt}
\end{equation}
in terms of the Riemannian connection $\Gamma^{\{\}}$ with  $D^{\{\}}\gamma=0$ and the {\em irreducible} piece
(\ref{axitor}) of the torsion.
Hence, in a RC spacetime a Dirac spinor does
only feel the {\em axial torsion} one--form $A$.  This can also be seen from
the  identity (3.6.13) of Ref. \cite{PRs} which  specializes here
to the ``on shell" commutation relation
\be
[\stackrel{\smile}{D}\, ,\stackrel{\smile}{D}]= \Omega^{\{\}} +
{i\over 4}\, \gamma_5 dA-  {i\over 8}\, m^2\sigma\,.  \label{comrel}
\ee
In contrast to Ref. \cite{ChandiaZ}, Eq. (27), there arise in (\ref{comrel})
no tensor or vector pieces of the torsion,
because our operator $\stackrel{\smile}{D}$
in (\ref{tt}) is the only possible result from the Lagrangian
(\ref{decldirac}), which is  {\em Hermitian} as required by QFT.

{}From the  Dirac equation (\ref{tt})  and its adjoint one can readily
deduce  the well--known ``classical axial anomaly"
$ dj_5 = d ({1\over 3} \overline{\psi}\sigma\wedge\gamma \psi)=
2miP  =2m i\overline{\psi}\gamma_5\psi$
for {\em massive} Dirac fields also
in a RC spacetime. If we restore chiral symmetry in the limit
$m\rightarrow 0$, this leads to {\em classical conservation law} of the
axial current for massless Weyl spinors, or since $dj =0$,
equivalently, for the {\em chiral current}
$ j_\pm :={1\over 2}
\overline{\psi}(1 \pm\gamma_5)\,^*\gamma\psi =
\overline{\psi}_{\rm L,R}\,^*\gamma\psi_{\rm L,R}$.

The Einstein--Cartan--Dirac (ECD) theory of a gravitationally
coupled  spin
$\frac{1}{2}$ fermion field provides a  {\em dynamical}  understanding
of the axial anomaly on a classical (i.e., not quantized) level.
{}From  Einstein's equations
$-(1/2)\,
\eta_{{\alpha}{\beta}{\gamma}}\wedge R^{{\beta}{\gamma}}=
\ell^{2}\,\Sigma_{\alpha} $
and the purely algebraic {\em Cartan relation}
$-(1/2) \eta_{\alpha\beta\gamma}\wedge T^{\gamma}=\ell^{2}
\tau_{\alpha\beta} =
-(\ell^2/4)\,\eta_{\alpha\beta\gamma\delta}\,\overline{\Psi}
\gamma_5\gamma^{\delta}\Psi\eta^{\gamma} $
one finds \cite{MMM96,MieKr}
\begin{equation}
dj_5 \cong 4 dC_{\rm TT}
={2\over{\ell^2}} \left(T^\alpha\wedge
T_\alpha+R_{\alpha\beta}\wedge\vartheta^\alpha\wedge\vartheta^\beta\right)
\label{eq:classan}
\end{equation}
which establishes  a link
to the NY  four form \cite{NY}, but only for  {\em massive} fields
\cite{MieKr}.

However, if we restore chiral invariance for the Dirac fields
in the limit $m\rightarrow 0$, we  find
within the dynamical framework of ECD theory that the
NY  four--form tends to zero ``on shell", i.e.
$dC_{\rm TT}\cong (1/4) dj_5 \rightarrow 0$.

This is consistent with the fact that a
Weyl spinor does not couple to torsion at all, because  the remaining axial torsion
$A$ becomes a {\em lightlike} covector, i.e.
$A_\alpha A^\alpha\eta =A\wedge\,^* \!A \cong (\ell^4/4)\,^*j_5\wedge j_5 =0$.
Here we implicitly assume that the light-cone structure
 of the axial covector $\,^*j_5$ is not spoiled by quantum corrections, i.e. that no
 ``Lorentz anomaly" occurs as in $n=4k +2$ dimensions \cite{Leut}.

\section{Chiral anomaly in QFT}
When quantum field theory (QFT) is involved, other boundary terms
may arise in  the {\em chiral anomaly} due to the
non--conservation of the axial current,  cf. \cite{zum,Hir}.

Now, to approach the anomaly in the context of spacetime with torsion,
we will proceed by switching off the curvature and concentrate
on the last term in the decomposed Dirac Lagrangian  (\ref{decldirac}).

Then,  this term  can be
regarded  as an {\em external} axial covector $A$ (in view of (\ref{comrel})
without Lorentz or ``internal" indices)  coupled to
the axial current $j_5$ of the Dirac field in an {\em initially flat}
spacetime. By applying
the result (11--225) of Itzykson and Zuber \cite{IZ}, we find
that only the term
$dA\wedge dA$ arises in the
axial anomaly, but {\em not} the NY  type term
$d\,^*A\sim dC_{\rm TT}$ as was
recently claimed \cite{ChandiaZ}.
After switching on the curved spacetime of Riemannian geometry, we finally
obtain for the axial anomaly
\be
\langle dj_5\rangle=  2m \langle iP \rangle+ \frac{1}{24\pi^2}
\left[ Tr\!\left(\Omega^{\{\}}\wedge\Omega^{\{\}}\right)  -
\frac{1}{4} dA\!\wedge\!dA\right].\label{anom} \ee

Besides other perturbative methods as point-splitting,  there is
the further option  to use
{\em dimensional regularization}.
If one adopts the $\gamma_5$ scheme of Ref. \cite{DK},
one  immediately concludes that only the result
(\ref{anom}) can appear.
The only effect of the $\gamma_5$ problem is the replacement
of the usual trace by a non-cyclic linear functional.
The anomaly appears as the sole effect of this non-cyclicity.
There is no
room for other sources of non-cyclicity apart from the very fermion
loops which produce the result (\ref{anom}).
The whole effect of non-cyclicity is to have an operator $\Delta$,
$\Delta^2=0$, and the anomaly is in the image modulo
the kernel of $\Delta$, which summarizes the fact that in this
$\gamma_5$ scheme no other anomalous contributions are possible
beside (\ref{anom}).

But at this stage we have not discussed the
possibility of a contorted spacetime which {\em cannot} be {\em adiabatically}
deformed  to the torsion-free case. In such a case it has been argued
\cite{ChandiaZ} that the boundary
term $dC_{\rm TT}$ occurs,
multiplied by a factor $M^2$. This factor $M^2$ corresponds to a
regulator mass in a Fujikawa type approach.
For instance, in the heat kernel approach, the first nontrivial terms
\cite{Yuri1,Yajima}, which potentially
could  contribute to the axial anomaly,  read
\ba
 Tr(\gamma_5 K_2) &=& -d\,^*\! A\,, \qquad
 \qquad {\cal K}=\,^*\!\stackrel{\smile}{D}\wedge^*\!\stackrel{\smile}{D}{}^*\!A \label{ano2} \\
 Tr(\gamma_5 K_4) &=& {1\over 6}
\left[Tr\left(\Omega^{\{\}}\wedge  \Omega^{\{\}}\right) -
{1\over 4} dA\wedge dA +d{\cal K}\right].\nonumber
 \ea
However, there is an essential difference in the physical dimensionality
of the
terms $K_2$ and
$K_4$. Whereas in $n=4$ dimensions the Pontrjagin type term $K_4$ is
dimensionless,  the term
$K_2\sim  2\ell^2 dC_{\rm TT}$ carries dimensions.
It can be  consistently absorbed
in a counterterm, and thus {\em discarded} from the final result for the
anomaly.

This is also in agreement with the analysis in \cite{bell} where,
in the framework of string theory, the chiral anomaly
in the presence of torsion
had a smooth adiabatic limit to the case of vanishing torsion.

In contrast, in Ref. \cite{ChandiaZ} it is argued that such contributions
can be maintained by absorbing the divergent factor in a rescaled
coframe $\widetilde\vartheta^\alpha:= M\vartheta^\alpha$ and propose
to consider the Wigner--In\"on\"u contraction $M\rightarrow \infty$ in the de
Sitter gauge approach \cite{GH}, with
$M\ell$ fixed.

Apart from the fact that
this would change also the dimension of $\psi$, in
order to retain  the physical dimension $[\hbar]$ of the Dirac action,
there are several points
which seem unsatisfactory in such an argument:

1. As  the difference (\ref{CTT}) of two Pontrjagin
classes, the term   $dC_{\rm TT}$ is a topological invariant after all.
Now, it is actually
{\em not} this term
which appears as the torsion-dependent extra contribution to the
anomaly, but more precisely $-d\,^*\!A= 2
\ell^2 dC_{\rm TT}$. Thus, measuring its proportion in units of the
topological invariant $dC_{\rm TT}$, we find that it vanishes when we consider
the proposed limit $M\to\infty$, keeping  $M\ell$ constant.

2. Instead of rescaling the vierbein, it is consistent
to compensate the ill-defined term by a counterterm.
This implies
that consistently a renormalization condition can be imposed which
guarantees the anomaly to have the
value (\ref{anom}).
Even if one  renders this extra term finite by a rescaling as in Ref.
\cite{ChandiaZ}, one has to confront the fact that a
(finite) renormalization condition can be imposed
which settles the anomaly at this value.
Further, if one were to keep this extra finite term, it would
be undetermined, and is thus not related to the anomaly at all.
Also, on-shell renormalization conditions adjust the
wave function renormalization of the fermion propagator to have
unit residuum at the physical mass. Any rescaling of the tetrad
cannot dispense for  the fact that the NY term needs
renormalization by itself,
as it is proven by the very calculation of \cite{ChandiaZ}.

3. From a {\em renormalization group} point of view, it is the scaling of  the
coupling which determines
the scaling of the anomaly (regarded as a Green's function), a property which is
desperately needed to maintain the validity, e.g., of
the proof of the Adler--Bardeen theorem.
Or, to put it otherwise, an anomaly is stable against radiative corrections for
the reason that such
corrections are compensated by a renormalization of the coupling.
While, on the other hand, the topological invariant of
Ref.\cite{ChandiaZ} has no such property,
its interpretation as an anomaly seems dubious to us.
The only consistent way out is to impose a renormalization
condition
which adjust the finite value of the NY term to be zero, which is
patently stable under radiative corrections.

4. Finally, it is well-known that usually the appearance of a chiral
anomaly is deeply connected with the presence of a {\em conformal anomaly}
\cite{Salam,Ellis,Kae}. This makes sense: usually,
conformal invariance is lost due to the (dynamical) generation of a scale.
But this is the very mechanism which destroys chiral invariance as well.
Thus, one would expect any  argument to fail
 trying  to combine strict conformal
invariance with a chiral anomaly.
\section{Conclusions}
Our conclusion is that the NY  term $dC_{\rm TT}$ does NOT
contribute to the {\em chiral anomaly} in $n=4$ dimensions, neither classically nor in quantum
field theory, in sharp contrast to Ref. \cite{ChandiaZ}.
We once more stress the interrelation between the scale and chiral invariance
\cite{Salam,Ellis,Kae}.
Since renormalization amounts to a continous scale deformation,
only the Riemannian part of the Pontrjagin term contributes
to the topology of the chiral anomaly.

The result of Chandia et al. is very different in spirit from a
typical anomaly, where, in pQFT, the relevant Green's function is
UV-finite and only implicitly depends on the scale via the
coupling. This contrasts the fact that the  Chandia et al. term
would depend {\em explicitly} on that scale.

Since the  $A$ is {\em not} a gauge field, one can also legitimately
absorb the contribution from the axial torsion covector $A$ into the redefined
{\em gauge--invariant} physical current
$\widehat j_5:=
j_5 +(1/96\pi^2) A\wedge dA  + M^2 \,^*\!A$,
where the last term depends explicitly on the regulator mass $M$.
It may arise from the counterterm $M^2\, A\wedge^*\!A$ to
the Dirac  Lagrangian  (\ref{tt}).
Then, due to
\ba \langle d\widehat j_5\rangle&=&
\langle dj_5 \rangle+(1/96\pi^2) dA\wedge dA  + M^2 d\,^*\!A \\  \label{anomriem}
&=&  2m \langle iP \rangle+ (1/24\pi^2)
Tr\!\left(\Omega^{\{\}}\wedge  \Omega^{\{\}}\right),\nonumber
\ea
only the {\em Riemannian} contribution remains for the axial anomaly of
this new {\em physical} current. A consistent way to avoid regularization
problems for $M\rightarrow \infty$ is to assume that the  ``photon" $A$ is
 is transverse, i.e. $ d\,^*\!A=0$, which is just the
 vanishing of the NY term.

On would surmize that
in  $n=2$ dimensional models only the term $d\,^* \!A$ survives in
the heat kernel expansion (\ref{ano2}), since it then has the correct
dimensions. However, it is well-known \cite{PRs} that in 2D the axial
torsion $A$ vanishes identically.

In general, the Pontrjagin topological invariant, as an integral over a {\em closed}
four-form,  depends  also upon the second fundamental form  of
the embedding of the boundary  $\partial M$ into $M$. In  Ref.
\cite{PW99} this is generalized  to spaces
with torsion  supporting our view that the index shall be independent
of regulator masses, hence excluding  contributions from
the NY term.
Since the integral $\int_{M} d\,^* \!A\equiv \int_{\partial M}\,^* \!A$
vanishes  {\em identically} for
manifolds without boundary, the NY invariant would  occur only
for non-closed manifolds, anyhow.

A situation where torsion
is indeed realized in a {\em discontinous} manner arises
for the cosmic string solution  within
the EC theory  \cite{SO,A94,TO}. We have shown in detail in Ref.
\cite{MieKr} that
the NY  term (\ref{eq:NY}) {\em vanishes identically} for this
example of a  spinning cosmic
 string exhibiting a {\em torsion line defect}.
The instantons of Ref. \cite{ChandiaZ} with non-vanishing NY  term are clearly
detached from an Einstein-type dynamics, and also the recent analysis in
Ref. \cite{ChopinSoo} fails to substantiate the presence of the NY term.

\acknowledgments
We would like to thank Alfredo Mac\'{\i}as, Friedrich W. Hehl, and
Yuri Obukhov
for useful comments and  the referee
for pointing out Ref. \cite{PW99}.
This work was partially supported by  CONACyT, grant No. 28339E, and the
joint German--Mexican project DLR--Conacyt
 E130--2924 and MXI 009/98 INF.
One of us (D.K.) acknowledges support by a Heisenberg fellowship of the DFG
and thanks the I.H.E.S. (Bures-sur--Yvette) for hospitality. E.W.M. thanks Noelia
M\'endez C\'ordova for encouragement.


\end{document}